\documentclass[aps,pra,english,onecolumn,notitlepage,showpacs,nofootinbib]{revtex4-1}
\usepackage{mathptmx}
\usepackage[T1]{fontenc}
\usepackage[latin9]{inputenc}
\usepackage{color}
\usepackage{babel}
\usepackage{textcomp}
\usepackage{bm}
\usepackage{graphicx}
\usepackage{amsmath}
\usepackage{amsthm}
\usepackage{amssymb}
\usepackage{epstopdf}
\usepackage[unicode=true,pdfusetitle,
bookmarks=true,bookmarksnumbered=false,bookmarksopen=false,
breaklinks=true,pdfborder={0 0 0},pdfborderstyle={},backref=false,colorlinks=true]{hyperref}
\usepackage{amsmath, amsfonts, amssymb, amsthm, mathrsfs, amssymb, braket, bbm, bm,graphicx,times,txfonts,color,subfigure}
\usepackage{hyperref}
\usepackage[table]{xcolor}

\hypersetup{
urlcolor=blue, citecolor=blue, linkcolor=blue}
\makeatletter

\theoremstyle{plain}
\newtheorem{Theorem}{Theorem}
\newtheorem{Proposition}{Proposition}

\newtheorem{Lemma}{Lemma}

\makeatother
\begin{document}

\title{Hermitian Kirkwood-Dirac real operators for discrete Fourier transformations}
\author{Jianwei Xu}
\email{xxujianwei@nwafu.edu.cn}
\affiliation{College of Science, Northwest A$\&$F University, Yangling, Shaanxi 712100,
China}

\begin{abstract}

The Kirkwood-Dirac (KD) distribution is a quantum state representation that
relies on two chosen fixed orthonormal bases, or alternatively, on the
transition matrix of these two bases. In recent years, it has been
discovered that the KD distribution has numerous applications in quantum
information science. The presence of negative or nonreal KD distributions
may indicate certain quantum features or advantages. If the KD distribution
of a quantum state consists solely of positive or zero elements, the state
is called a KD positive state. Consequently, a crucial inquiry arises
regarding the determination of whether a quantum state is KD positive when
subjected to various physically relevant transition matrices. When the
transition matrix is discrete Fourier transform (DFT) matrix of dimension $p$
[\href{https://doi.org/10.1063/5.0164672}%
{J. Math. Phys. 65, 072201 (2024)}] or $p^{2}$ [\href{https://dx.doi.org/10.1088/1751-8121/ad819a}%
{J. Phys. A: Math. Theor. 57 435303 (2024)}] with $p$ being prime, it is proved that any KD
positive state can be expressed as a convex combination of pure KD positive
states. In this work, we prove that when the transition matrix is the DFT
matrix of any finite dimension, any KD positive state can be expressed as a
real linear combination of pure KD positive states.
\end{abstract}

\maketitle

\section{Introduction and main results}
\setcounter{equation}{0} \renewcommand%
\theequation{1.\arabic{equation}}

The Kirkwood-Dirac (KD) distribution, originally introduced by Kirkwood
\cite{Kirkwood-1933-PR} and Dirac \cite{Dirac-1945-RMP}, is a representation of quantum states. In
recent years, KD distribution has found many applications in quantum
information science \cite{NYH-2021-JPA,Bievre-2024-NJP,Bievre-2021-PRL,Quantum-2022-Lostaglio,Bievre-2023-JMP,Budiyono-2023-PRA,
Budiyono-2023-JPA,PA-Rastegin-2023,Wagner-2024-QST,Bievre-JMP-2024,Budiyono-2024-PRA,Selby-2024-PRA,
Fu-2024-PRA,Fan-2024-PS,Xu-2024-PLA,Yang-2024-JPA,Bievre-202405-arxiv,Bievre-202407-arxiv}, we refer to Refs. \cite{NYH-2021-JPA} and \cite{Bievre-2024-NJP} for recent reviews on the
KD distribution, diverse applications, and some theoretical advances.

According to the postulates of quantum mechanics, a quantum system is
described by a complex Hilbert space \cite{Nielsen-2010-book}. We denote the specific complex Hilbert
space under consideration by $\boldsymbol{H}_{d}^{C}$ with $d$ the dimension
of $\boldsymbol{H}_{d}^{C}.$ We fix two orthonormal bases of $\boldsymbol{H}%
_{d}^{C}$, $A=\{|a_{j}\rangle \}_{j=1}^{d}$ and $B=\{|b_{k}\rangle
\}_{k=1}^{d}$. A quantum state is mathematically described by a density
operator $\rho $ on $\boldsymbol{H}_{d}^{C},$ that is, $\rho $ is positive
semidefinite and of unit trace. The KD distribution of $\rho $ with respect
to bases $A$ and $B$ is defined as
\begin{eqnarray}
Q_{jk}=\langle a_{j}|\rho |b_{k}\rangle \langle b_{k}|a_{j}\rangle , \ \forall
(j,k)\in \llbracket{1,d}\rrbracket^{2},     \label{eq1.1}
\end{eqnarray}
where $\llbracket{l,l+m}\rrbracket$ denote the set of consecutive integers from $l$ to $l+m,$
for example, $\llbracket{2,5}\rrbracket=\{2,3,4,5\}$. It is easily checked that
\begin{eqnarray}
\sum_{j=1}^{d}Q_{jk}=\langle b_{k}|\rho |b_{k}\rangle,
\ \sum_{k=1}^{d}Q_{jk}=\langle a_{j}|\rho |a_{j}\rangle,
\ \sum_{j,k=1}^{d}Q_{jk}=1,
\ \ \forall (j,k)\in \llbracket{1,d}\rrbracket^{2}.   \label{eq1.2}
\end{eqnarray}%
This shows that the marginal distributions $\{\sum_{j=1}^{d}Q_{jk}%
\}_{k=1}^{d}$ and $\{\sum_{k=1}^{d}Q_{jk}\}_{j=1}^{d}$ of the KD
distribution $\{Q_{jk}\}_{j,k=1}^{d}$ are all probability distributions.
Since there may be some $Q_{jk}$ negative or complex, we call $%
\{Q_{jk}\}_{j,k=1}^{d}$ a quasiprobability distribution. If $%
\{Q_{jk}\}_{j,k=1}^{d}$ is a true probability distribution, i.e., $%
Q_{jk}\geq 0$ for all $(j,k)\in \llbracket{1,d}\rrbracket^{2},$ we say that $%
\rho $ is a KD positive state with respect to $A$ and $B.$

The KD distribution has found applications in various scenarios of quantum
information science, where the presence of KD nonpositivity may exhibit
quantum advantages. Consequently, an essential theoretical question arises:
how can we ascertain whether a quantum state is KD positive? Till now there
have been limited findings on this topic \cite{NYH-2021-JPA,Bievre-2021-PRL,Bievre-2023-JMP,Xu-2024-PLA,Bievre-JMP-2024,
Bievre-JMP-2024,Bievre-202405-arxiv,Bievre-202407-arxiv,Yang-2024-JPA}. Some conditions for pure
KD positive states were established \cite{NYH-2021-JPA,Bievre-2021-PRL,Bievre-2023-JMP,Xu-2024-PLA}. Conditions for mixed KD positive states are more complicated in general than pure states \cite{Bievre-JMP-2024,Bievre-JMP-2024,Bievre-202405-arxiv,Bievre-202407-arxiv,Yang-2024-JPA}.

KD distribution depends on the fixed orthonormal bases $A=\{|a_{j}\rangle
\}_{j=1}^{d}$ and $B=\{|b_{k}\rangle \}_{k=1}^{d},$ or, in other words, on
the transition matrix $U$ with the elements $U_{jk}$ defined by $A$ and $B$
as
\begin{eqnarray}
U_{jk}=\langle a_{j}|b_{k}\rangle , \ \forall (j,k)\in \llbracket{1,d}\rrbracket^{2}.     \label{eq1.3}
\end{eqnarray}
One can check that $U$ is unitary, i.e., $U^{\dagger }U=I_{d}.$ We use $%
U^{T},$ $U^{\ast }$ and $U^{\dagger }$ to denote the transpose, complex
conjugate and transpose conjugate (or called Hermitian conjugate) of $U,$
use $I_{d}$ to denote the identity operator of dimension $d.$ In principle,
there is an infinite number of transition matrices, in practice we are only
interested in some transition matrices which are important in applications.
Among them, discrete Fourier transform (DFT) is widely used in both
mathematics and physics. The DFT matrix $U$ has the elements
\begin{eqnarray}
U_{jk}=\langle a_{j}|b_{k}\rangle =\frac{1}{\sqrt{d}}\omega _{d}^{jk},
\ \ \omega
_{d}=e^{i\frac{2\pi }{d}},
\ \ i=\sqrt{-1},
\ \ \forall (j,k)\in \llbracket{1,d}\rrbracket^{2}.    \label{eq1.4}
\end{eqnarray}%
In this paper we always assume that the two orthonormal bases $A=\{|a_{j}\rangle
\}_{j=1}^{d}$ and $B=\{|b_{k}\rangle \}_{k=1}^{d}$ are fixed, and the transition matrix
is DFT matrix.

For DFT of general dimension $d$, it is shown that a pure state is KD
positive if and only if the pure state can be expressed as \cite{Bievre-2021-PRL,Xu-2024-PLA}
\begin{eqnarray}
|\psi _{d_{1},d_{2},j,k}\rangle =\frac{e^{i\alpha }}{\sqrt{d_{1}}}%
\sum_{j^{\prime }=1}^{d_{1}}\omega _{d_{1}}^{jj^{\prime }}|a_{k+j^{\prime
}d_{2}}\rangle =\frac{e^{i\alpha }}{\sqrt{d_{2}}}\omega
_{d}^{-jk}\sum_{k^{\prime }=1}^{d_{2}}\omega _{d_{2}}^{-kk^{\prime
}}|b_{j+k^{\prime }d_{1}}\rangle    \label{eq1.5}
\end{eqnarray}%
where $d_{1}\in\mathbb{N}_{+},$ $d=d_{1}d_{2},$ $\alpha \in\mathbb{R}
,$ $j\in \llbracket{1,d_{1}}\rrbracket,$ $k\in \llbracket{1,d_{2}}\rrbracket,$ $%
\mathbb{N}_{+}$ denotes the set of positive integers, $\mathbb{R}
$ denotes the set of real numbers. We use $\mathcal{E}_{\text{KD+}}^{\text{%
pure}}$ to denote the set of all pure KD positive states for DFT of
dimension $d,$
\begin{eqnarray}
\mathcal{E}_{\text{KD+}}^{\text{pure}}=\{|\psi _{d_{1},d_{2},j,k}\rangle
\langle \psi _{d_{1},d_{2},j,k}|:d_{1}\in
\mathbb{N}
_{+},d=d_{1}d_{2},j\in \llbracket{1,d_{1}}\rrbracket, k\in \llbracket{1,d_{2}}\rrbracket\}. \label{eq1.6}
\end{eqnarray}%
Also, let $\mathcal{E}_{\text{KD+}}$ denote all KD positive states for DFT
of dimension $d,$ conv($\mathcal{E}_{\text{KD+}}^{\text{pure}}$) denote the
set of convex combinations of the elements in set $\mathcal{E}_{\text{KD+}}^{%
\text{pure}},$ span$_{\mathbb{R}}(\mathcal{E}_{\text{KD+}}^{\text{pure}}$) denote the real vector space
linearly spanned by $\mathcal{E}_{\text{KD+}}^{\text{pure}}$ over $\mathbb{R}.$

When $d=p$ being prime, Eq. (\ref{eq1.6}) yields that the KD positive pure states are just the basis states,
\begin{eqnarray}
\mathcal{E}_{\text{KD+}}^{\text{pure}}=\{|a_{j}\rangle \langle
a_{j}|\}_{j=1}^{d}\cup \{|b_{k}\rangle \langle b_{k}|\}_{j=1}^{d}.  \label{eq1.7}
\end{eqnarray}%
When $d=p$ being prime, Langrenez, Arvidsson-Shukur, and De Bi\`evre proved
that any KD positive state $\rho $ can be
expressed as a convex combination of pure KD positive states  \cite{Bievre-JMP-2024},
\begin{eqnarray}
\rho =\sum_{j=1}^{d}p_{j}|a_{j}\rangle \langle
a_{j}|+\sum_{k=1}^{d}q_{k}|b_{k}\rangle \langle b_{k}|,  \label{eq1.8}
\end{eqnarray}%
with $\{p_{j}\}_{j=1}^{d}\cup \{q_{k}\}_{k=1}^{d}$ is a probability
distribution. This result finds out all KD positive states for DFT of prime
dimension. Eq. (\ref{eq1.8}) says that
\begin{eqnarray}
\mathcal{E}_{\text{KD+}}=\text{conv}(\mathcal{E}_{\text{KD+}}^{\text{pure}})  \label{eq1.9}
\end{eqnarray}%
when $d=p$ is prime. Yang et.al. proved that Eq. (\ref{eq1.9}) also holds for $d=p^{2}$
with $p$ being prime \cite{Yang-2024-JPA}. Further, Yang et.al. conjectured that Eq. (\ref{eq1.9}) holds for
any dimension $d$ \cite{Yang-2024-JPA}.

In this work, we prove Theorem \ref{Theorem-1} below, Theorem \ref{Theorem-1} can be seen as a weaker
result of Eq. (\ref{eq1.9}). To state Theorem \ref{Theorem-1}, we introduce the real vector space $V_{%
\text{KDr}}$ and the gcd-sum function $\gamma(d)$. Let $V_{\text{KDr}}$ denote
the set of all Hermitian (also called self-adjoint) operators on $\boldsymbol{H}_{d}^{C}$ with real KD distribution, that is,
\begin{eqnarray}
V_{\text{KDr}}=\{F:F=F^{\dagger },Q_{jk}\in\mathbb{R}
\text{ for all }(j,k)\in \llbracket{1,d}\rrbracket^{2}\}.  \label{eq1.10}
\end{eqnarray}%
We can check that $V_{\text{KDr}}$ is a real vector space. Note that for any
linear operator $O$ on $\boldsymbol{H}_{d}^{C},$ the KD distribution $Q_{jk}$
of $O$ can be similarly defined as in Eq. (\ref{eq1.1}) by replacing $\rho $ with $O.$
We call $O$ a KD positive (real) operator if $Q_{jk}\geq 0$ ($Q_{jk}\in
\mathbb{R}$) for all $(j,k)\in \llbracket{1,d}\rrbracket^{2}.$
Suppose the standard prime factorization of $d$ is (see for example Ref. \cite{Rosen-2011-book})
\begin{eqnarray}
d=p_{1}^{\alpha _{1}}p_{2}^{\alpha _{2}}...p_{s}^{\alpha _{s}},  \label{eq1.11}
\end{eqnarray}%
where $\{s,\alpha _{1},\alpha _{2},...,\alpha _{s}\}\subseteq
\mathbb{N}_{+},$ $\{p_{1},p_{2},...,p_{s}\}$ are all prime, $p_{1}<p_{2}<...<p_{s}.$  For Eq. (\ref{eq1.11}), the gcd-sum
function (called also Pillai's arithmetical function) $\gamma (d)$ is
defined as
\begin{eqnarray}
\gamma (d)=\sum_{n=1}^{d}\text{gcd}(n,d)=d\Pi _{j=1}^{s}\left[1+\alpha _{j}\left(1-%
\frac{1}{p_{j}}\right)\right],   \label{eq1.12}
\end{eqnarray}%
where gcd stands for greatest common divisor. We provide a proof for the
second equality of Eq. (\ref{eq1.12}) in \hyperlink{Appendix}{Appendix}. With such notation, we state Theorem 1 as follows.
\begin{Theorem} \label{Theorem-1}
For DFT of general dimension $d,$ we have
\begin{eqnarray}
\text{dim}\left( V_{\text{KDr}}\right)  &=&\gamma (d), \label{eq1.13} \\
V_{\text{KDr}} &=&\text{span}_{\mathbb{R}
}(\mathcal{E}_{\text{KD+}}^{\text{pure}}).  \label{eq1.14}
\end{eqnarray}
\end{Theorem}
Theorem 1 implies that for DFT of general dimension $d,$ any KD real
(positive) state $\rho $ can be expressed as a real linear combination of pure KD
positive states, i.e.,
\begin{eqnarray}
\rho =\sum_{d=d_{1}d_{2},j\in \llbracket{1,d_{1}}\rrbracket,k\in \llbracket{1,d_{2}}\rrbracket}\lambda _{d_{1},d_{2},j,k}|\psi _{d_{1},d_{2},j,k}\rangle \langle\psi _{d_{1},d_{2},j,k}|,  \label{eq1.15}
\end{eqnarray}%
with $\lambda _{d_{1},d_{2},j,k}\in
\mathbb{R}
,$ $\sum_{d=d_{1}d_{2},j\in \llbracket{1,d_{1}}\rrbracket,k\in \llbracket{1,d_{2}}\rrbracket}\lambda _{d_{1},d_{2},j,k}=1.$

The rest of this paper is to prove Theorem 1. In section \ref{Section-2}, we prove Eq. (\ref{eq1.13}).
In section \ref{Section-3} we give an example of Eq. (\ref{eq1.13}) for dimension 6. In section \ref{Section-4}, we prove Eq. (\ref{eq1.14}). Section \ref{Section-5} is a short summary. In \hyperlink{Appendix}{Appendix} we provide some necessary details.

\section{Proof of dim$\left( V_{\text{KDr}}\right) =\gamma (d)$ in Eq. (\ref{eq1.13})}    \label{Section-2}
\setcounter{equation}{0} \renewcommand%
\theequation{2.\arabic{equation}}

We first prove Lemma \ref{Lemma-1} below. Lemma \ref{Lemma-1} establishes a necessary and sufficient
condition for a Hermitian operator $F$ being KD real. For DFT of general
dimension $d,$ suppose the operator $F$ is Hermitian, that is, $F=F^{\dag
}.$ We express $F$ in the orthonormal basis $A=\{|a_{j}\rangle \}_{j=1}^{d}$
as
\begin{eqnarray}
F=\sum_{j,k=1}^{d}|a_{j}\rangle \langle a_{j}|F|a_{k}\rangle \langle
a_{k}|=\sum_{j,k=1}^{d}F_{jk}^{A}|a_{j}\rangle \langle a_{k}|,   \label{eq2.1}
\end{eqnarray}
with $F_{jk}^{A}=\langle a_{j}|F|a_{k}\rangle .$ We can regard $F^{A}$ as a
matrix with entries $F^{A}=\{F_{jk}^{A}\}_{j,k=1}^{d}.$ Let $%
\mathbb{Z}
$ represent the set of all integers, $\mathbb{Z}_{d}$ represent the set of all integers modulo $d$, we can take $\mathbb{Z}_{d}=\llbracket{1,d}\rrbracket$ or $\mathbb{Z}
_{d}=\llbracket{0,d-1}\rrbracket$ without ambiguity. For convenience of expressing indexes $%
j\in \llbracket{1,d}\rrbracket,$ we let $j\in
\mathbb{Z}$, and regard $j$ as $k$ in the sense $j\equiv k($mod $d)$ with $k\in
\mathbb{Z}
_{d}.$ We often abbreviate $j\equiv l($mod $d)$ as $j\equiv l$ omitting (mod
$d$)$,$ for example, $d\equiv 0$, $-1\equiv d-1$ and $n\equiv d-n.$
\begin{Lemma} \label{Lemma-1}
For DFT of general dimension $d,$ suppose the operator $F$ is
Hermitian, then
\begin{eqnarray}
F\in V_{\text{KDr}}\Longleftrightarrow F_{j,k}^{A}=F_{2j-k,j}^{A} \ \forall (j,k)\in \llbracket{1,d}\rrbracket^{2},  \label{eq2.2}
\end{eqnarray}
where $F_{j,k}^{A}=\langle a_{j}|F|a_{k}\rangle .$
\end{Lemma}
\textbf{Proof of Lemma 1.} For the matrices $Q=(Q_{jk})_{j,k=1}^{d},$ $%
U=(U_{jk})_{j,k=1}^{d}$ and $F=(F_{j,k}^{A})_{j,k=1}^{d},$  we express the KD distribution in Eq. (\ref{eq1.1}) in the matrix form as in Ref. \cite{Fan-2024-PS},
\begin{eqnarray}
Q=(F^{A}U)\circ U^{\ast },   \label{eq2.3}
\end{eqnarray}%
where $\circ $ stands for the Hadamard product (entrywise product) of
matrices. $F\in V_{\text{KDr}}$ implies $Q=Q^{\ast },$ then
\begin{eqnarray}
(F^{A}U)\circ U^{\ast } &=&[(F^{A})^{\ast }U^{\ast }]\circ U,  \nonumber  \\
F^{A} &=&\{[(F^{A})^{\ast }U^{\ast }]\circ U\circ U^{\ast (-1)}\}U^{\dagger},  \label{eq2.4}
\end{eqnarray}%
where $U^{\ast (-1)}$ denotes the matrix obtained by taking the reciprocal
of each entry of $U^{\ast },$ that is, $U^{\ast (-1)}=(\frac{1}{U_{jk}^{\ast
}})_{j,k=1}^{d}.$ For DFT matrix defined in Eq. (\ref{eq1.3}), taking the facts $U_{jk}=%
\frac{1}{\sqrt{d}}\omega _{d}^{jk},$ $U_{jk}^{\ast }=$ $\frac{1}{\sqrt{d}}%
\omega _{d}^{-jk},$ $U=U^{T},$ $U^{\dagger }=U^{\ast }$ and $U_{jk}^{\ast
(-1)}=\sqrt{d}\omega _{d}^{jk}$ into Eq. (\ref{eq2.4}), we have%
\begin{eqnarray*}
F_{j,k}^{A} &=&\sum_{l=1}^{d}\{[(F^{A})^{\ast }U^{\ast }]\circ U\circ
U^{\ast (-1)}\}_{jl}U_{lk}^{\dagger } \\
&=&\sum_{l=1}^{d}[(F^{A})^{\ast }U^{\ast }]_{jl}U_{jl}U_{jl}^{\ast
(-1)}U_{lk}^{\dagger } \\
&=&\sum_{l,m=1}^{d}(F^{A})_{jm}^{\ast }U_{ml}^{\ast }U_{jl}U_{jl}^{\ast
(-1)}U_{lk}^{\dagger } \\
&=&\frac{1}{d}\sum_{l,m=1}^{d}F_{mj}^{A}\omega _{d}^{-ml}\omega
_{d}^{jl}\omega _{d}^{jl}\omega _{d}^{-kl} \\
&=&\frac{1}{d}\sum_{l,m=1}^{d}F_{mj}^{A}\omega _{d}^{l(2j-m-k)} \\
&=&\sum_{m=1}^{d}F_{mj}^{A}\left[ \frac{1}{d}\sum_{l=1}^{d}\omega
_{d}^{l(2j-m-k)}\right] \\
&=&\sum_{m=1}^{d}F_{mj}^{A}\delta _{2j-m-k,0}=F_{2j-k,j}^{A}.
\end{eqnarray*}
In above derivation, we used $\frac{1}{d}\sum_{k=1}^{d}\omega _{d}^{jk}=\delta _{jd},$
here $\delta _{jd}=1$ if $j=d$, $\delta _{jd}=0$ if $j\neq d.$ We
also used $(F^{A})_{jm}^{\ast }=F_{mj}^{A}$ since $F=F^{\dag }.$ We then
proved Lemma 1. $\hfill\blacksquare$

Remark that, when $d=p$ being prime Lemma \ref{Lemma-1} was proved in Ref. \cite{Bievre-JMP-2024}, for
general $d$ Lemma \ref{Lemma-1} was also proved in Ref. \cite{Yang-2024-JPA}. Our proof of Lemma 1 is different
from the proofs in Refs. \cite{Bievre-JMP-2024,Yang-2024-JPA}.

Now we prove dim$\left( V_{\text{KDr}}\right) =\gamma (d)$ in Eq. (\ref{eq1.13}).
We use Lemma \ref{Lemma-1} to study dim$(V_{\text{KDr}}).$ Eq. (\ref{eq2.2}) implies
\begin{eqnarray}
F_{j,k}^{A}=F_{j+(j-k),k+(j-k)}^{A}=F_{j+2(j-k),k+2(j-k)}^{A}...=F_{j+l(j-k),k+l(j-k)}^{A}=...,\ \forall (j,k,l)\in \llbracket{1,d}\rrbracket^{3}.   \label{eq2.5}
\end{eqnarray}
Eq. (\ref{eq2.5}) makes the entries $\{F_{j+l(j-k),k+l(j-k)}^{A}\}_{l=1}^{d}$ equal to
each other.

We decompose $F^{A}=(F_{jk}^{A})_{j,k=1}^{d}$ into
\begin{eqnarray}
\{F_{jk}^{A}\}_{j,k=1}^{d}=\cup _{n=0}^{d-1}F_{n}^{A}, \ \
F_{n}^{A}=\{F_{jk}^{A}\}_{j-k\equiv n}.   \label{eq2.6}
\end{eqnarray}
Let $|F_{n}^{A}|$ represent the number of elements in the set $F_{n}^{A}.$
Clearly, $|F_{n}^{A}|=d.$ Using $j-k\equiv n,$ $\{F_{j+l(j-k),k+l(j-k)}^{A}%
\}_{l=1}^{d}$ can be recast as
\begin{eqnarray}
F_{(k+n,k)}^{A}=\{F_{k+(m+1)n,k+mn}^{A}:m\in \llbracket{1,d}\rrbracket\}.   \label{eq2.7}
\end{eqnarray}
Note that $F_{(k+n,k)}^{A}$ does not must contain $d$ elements because there
may be some entries appeared repeatedly. As pointed out in Ref.
\cite{Yang-2024-JPA} that $|F_{(k+n,k)}^{A}|=\frac{d}{\text{gcd}(n,d)}.$
Consequently, Eq. (\ref{eq2.7}) can also be written as
\begin{eqnarray}
F_{(k+n,k)}^{A}=\left\{F_{k+(m+1)\text{gcd}(n,d),k+m\text{gcd}(n,d)}^{A}:m\in
\left\llbracket{1,\frac{d}{\text{gcd}(n,d)}}\right\rrbracket\right\}.   \label{eq2.8}
\end{eqnarray}

We can further decompose $F_{n}^{A}=\{F_{jk}^{A}\}_{j-k\equiv n}$ into
\begin{eqnarray}
F_{n}^{A}=\cup _{k=1}^{\text{gcd}(n,d)}F_{(k+n,k)}^{A}.   \label{eq2.9}
\end{eqnarray}

Eqs. (\ref{eq2.6},\ref{eq2.8},\ref{eq2.9}) lead to a partition of $\{F_{jk}^{A}\}_{j,k=1}^{d},$ that is, $%
\{F_{jk}^{A}\}_{j,k=1}^{d}$ are divided into $\gamma (d)=\sum_{n=1}^{d}$gcd$%
(n,d)$ classes.

From $F=F^{\dagger }$ and Eq. (\ref{eq2.6}) we see that for any $F_{(k+n,k)}^{A}$ we
have $F_{n}^{A}\ni F_{k+n,k}^{A}=(F_{k,k+n}^{A})^{\ast }\in
(F_{-n}^{A})^{\ast },$ here $(F_{n}^{A})^{\ast }=\{(F_{jk}^{A})^{\ast
}\}_{j-k\equiv n}.$ Since $|F_{(k+n,k)}^{A}|=\frac{d}{\text{gcd}(n,d)}$ and
gcd$(n,d)=$gcd$(-n,d),$ then
\begin{eqnarray}
F_{(k+n,k)}^{A}=(F_{(k-n,k)}^{A})^{\ast }.   \label{eq2.10}
\end{eqnarray}

If $F_{(k+n,k)}^{A}=F_{(k-n,k)}^{A},$ then $n\equiv -n$ (mod $d$), this
requires $n=0$ or $n=d/2$ (this happens only when $2|d,$ that is, $2$
divides $d$). $F_{(k+n,k)}^{A}=F_{(k-n,k)}^{A}$ implies $F_{(j,j+n)}^{A}\subseteq\mathbb{R}.$

\hypertarget{ii.1}{}
(ii.1). When $n=0,$ $F_{(k,k)}^{A}=\{F_{k,k}^{A}\}\subseteq\mathbb{R}.$   

\hypertarget{ii.2}{}
(ii.2). When $n=d/2,$ $F_{(k+d/2,k)}^{A}=\{F_{k+d/2,k}^{A},F_{k,k+d/2}^{A}\}\subseteq\mathbb{R}.$      

Combining Eqs. (\ref{eq2.6},\ref{eq2.8},\ref{eq2.9},\ref{eq2.10}), we compute dim$(V_{\text{KDr}}).$ From (\hyperlink{ii.1}{ii.1}), we see that
$F_{0}^{A}=\cup _{k=1}^{d}\{F_{k,k}^{A}\}$ with $\{F_{k,k}^{A}\}_{j=1}^{d}\subseteq\mathbb{R},$ then
\begin{eqnarray}
\dim (F_{0}^{A})=d=|F_{0}^{A}|=\gcd (0,d).   \label{eq2.11}
\end{eqnarray}%
From (\hyperlink{ii.2}{ii.2}), we see that $F_{d/2}^{A}=\cup _{k=1}^{d/2}F_{(k+d/2,k)}^{A}$
with $F_{d/2}^{A}\subseteq\mathbb{R},$ then
\begin{eqnarray}
\dim (F_{d/2}^{A})=d/2=|F_{d/2}^{A}|=\gcd (d/2,d).  \label{eq2.12}
\end{eqnarray}%
For any $F_{k+n,k}^{A}\in F_{(k+n,k)}^{A}$, if $n\not\equiv 0$ and $n\not\equiv \frac{d}{2},$ we write $F_{k+n,k}^{A}$ as
\begin{eqnarray}
F_{k+n,k}^{A}=\text{Re}F_{k+n,k}^{A}+i\text{Im}F_{k+n,k}^{A},  \label{eq2.13}
\end{eqnarray}
then Eq. (\ref{eq2.10}) implies that any $F_{k-n,k}^{A}\in F_{(k-n,k)}^{A}$ must be%
\begin{eqnarray}
F_{k-n,k}^{A}=\text{Re}F_{k+n,k}^{A}-i\text{Im}F_{k+n,k}^{A}.  \label{eq2.14}
\end{eqnarray}
As a result,
\begin{eqnarray}
&\dim(F_{(k+n,k)}^{A}\cup F_{(k-n,k)}^{A})=2,  \label{eq2.15} \\
&\dim (F_{n}^{A}\cup F_{-n}^{A})=2\gcd (n,d).   \label{eq2.16}
\end{eqnarray}

Eqs. (\ref{eq2.9},\ref{eq2.11},\ref{eq2.12},\ref{eq2.16}) together implies Eq. (\ref{eq1.13}). We then proved Eq. (\ref{eq1.13}).

Remark that, when $d=p$ being prime, Eqs. (\ref{eq1.12},\ref{eq1.13}) yield $\dim (V_{\text{KDr}%
})=2p-1,$ this returns to the same result in Ref. \cite{Bievre-JMP-2024}; when $d=p^{2}$ with
$p$ being prime, Eqs. (\ref{eq1.12},\ref{eq1.13}) yields $\dim (V_{\text{KDr}})=3p^{2}-2p,$ this
returns to the same result in Ref. \cite{Yang-2024-JPA}; when $d=pq$ with $p$ and $q$ both
being prime, Eqs. (\ref{eq1.12},\ref{eq1.13}) yield $\dim (V_{\text{KDr}})=(2p-1)(2q-1),$ this
returns to the same result in Ref. \cite{Yang-2024-JPA}.

For the gcd-sum function, we have Proposition 1 below.
\begin{Proposition}    \label{Proposition-1}
Suppose $\gamma (d)$ is the gcd-sum function of $d,$ $\tau (d)
$ is the number of positive divisors of $d,$ $\mu (d)$ is the M\"{o}bius
function of $d.$ Then we have
\begin{eqnarray}
\Gamma (d) &=&\sum_{j|d,j\in \llbracket{1,d}\rrbracket}\gamma (j)=d\tau (d),  \label{eq2.17} \\
\gamma (d) &=&\sum_{j|d,j\in \llbracket{1,d}\rrbracket}j\tau (j)\mu \left(\frac{d}{j}\right).  \label{eq2.18}
\end{eqnarray}
\end{Proposition}

We provide a proof for Proposition 1 in \hyperlink{Appendix}{Appendix}. From Eqs. (\ref{eq1.5},\ref{eq1.6}) we see that $%
d\tau (d)$ is just the number of all pure KD positive states, i.e.,
\begin{eqnarray}
|\mathcal{E}_{\text{KD+}}^{\text{pure}}|=d\tau (d).   \label{eq2.19}
\end{eqnarray}%
Together with $\dim (V_{\text{KDr}})=\gamma (d),$ thus Proposition 1
established a relation between $V_{\text{KDr}}$ and $\mathcal{E}_{\text{KD+}%
}^{\text{pure}}.$

\section{An example: \ $\text{dim}(V_{\text{KDr}})$ for $d=6$}    \label{Section-3}
\setcounter{equation}{0} \renewcommand%
\theequation{3.\arabic{equation}}
As a demonstration of the structure of $V_{\text{KDr}}$ described in Eqs. (\ref{eq2.5}-\ref{eq2.16}), we consider Example 1 of $d=6$ below. This example was also previously discussed in Ref. \cite{Yang-2024-JPA}.

\textbf{Example 1.} When $d=6,$ $F\in V_{\text{KDr}},$ we write $F^{A}=(F_{jk}^{A})_{j,k=1}^{6}$ in the matrix form as
\begin{eqnarray}
\begin{array}{c}
\cellcolor{red!50}F_{0}^{A} \\
\cellcolor{green!100}F_{1}^{A} \\
\cellcolor{yellow!100}F_{2}^{A} \\
\cellcolor{blue!50}F_{3}^{A} \\
\cellcolor{yellow!100}F_{4}^{A} \\
\cellcolor{green!100}F_{5}^{A} \\
\\
\\
\\
\\
\\
\\
\end{array}%
\left[
\begin{array}{cccccc}
&  &  &  &  &  \\
\cellcolor{red!50}_{(1)}\rho _{11} & \rho _{12} & \rho _{13} & \rho _{14} &
\rho _{15} & \rho _{16} \\
\cellcolor{green!100}_{(7)}\rho _{21} & \cellcolor{red!50}_{(2)}\rho _{22} & \rho _{23} & \rho _{24} &\rho _{25} &\rho _{26} \\
\cellcolor{yellow!100}_{(9)}\rho _{31} & \cellcolor{green!100}_{(7)}\rho _{32} & \cellcolor{red!50}_{(3)}\rho _{33} & \rho _{34} & \rho_{35} & \rho _{36} \\
\cellcolor{blue!50}_{(13)}\rho _{41} & \cellcolor{yellow!100}_{(10)}\rho _{42} & \cellcolor{green!100}_{(7)}\rho _{43} & \cellcolor{red!50}_{(4)}\rho _{44} & \rho _{45} &\rho _{46} \\
\cellcolor{yellow!100}_{(11)}\rho _{51} & \cellcolor{blue!50}_{(14)}\rho _{52} & \cellcolor{yellow!100}_{(9)}\rho _{53} & \cellcolor{green!100}_{(7)}\rho _{54} & \cellcolor{red!50}_{(5)}\rho _{55} &\rho _{56} \\
\cellcolor{green!100}_{(8)}\rho _{61} & \cellcolor{yellow!100}_{(12)}\rho _{62} & \cellcolor{blue!50}_{(15)}\rho _{63} & \cellcolor{yellow!100}_{(10)}\rho _{64} & \cellcolor{green!100}_{(7)}\rho _{65} & \cellcolor{red!50}_{(6)}\rho
_{66} \\
& \cellcolor{green!100}_{(8)}\rho _{12} & \cellcolor{yellow!100}_{(11)}\rho _{13} & \cellcolor{blue!50}_{(13)}\rho _{14} & \cellcolor{yellow!100}_{(9)}\rho _{15}
&\cellcolor{green!100} _{(7)}\rho _{16} \\
&  & \cellcolor{green!100}_{(8)}\rho _{23} & \cellcolor{yellow!100}_{(12)}\rho _{24} & \cellcolor{blue!50}_{(14)}\rho _{25} & \cellcolor{yellow!100}_{(10)}\rho_{26} \\
&  &  & \cellcolor{green!100}_{(8)}\rho _{34} & \cellcolor{yellow!100}_{(11)}\rho _{35} & \cellcolor{blue!50}_{(15)}\rho _{36} \\
&  &  &  & \cellcolor{green!100}_{(8)}\rho _{45} & \cellcolor{yellow!100}_{(12)}\rho _{46} \\
&  &  &  &  & \cellcolor{green!100}_{(8)}\rho _{56}%
\end{array}%
\right]  \label{eq3.1}
\end{eqnarray}
where we added 5 rows to the bottom of matrix to repeat the entries $\{F_{jk}^{A}\}_{1\leq j<k\leq 6}$ and added a column to the left side of matrix to express $\{F_{n}^{A}\}_{n=0}^{5}$.

(iii.0). When $n\equiv j-k\equiv 0,$ gcd$(0,6)=$gcd$(6,6)=6$, $\frac{6}{\text{%
gcd}(6,6)}=1.$ Eqs. (\ref{eq2.6},\ref{eq2.8}) yield
\begin{eqnarray}
F_{0}^{A}
&=&\{F_{11}^{A},F_{22}^{A},F_{33}^{A},F_{44}^{A},F_{55}^{A},F_{66}^{A}\}; \\
F_{(k,k)}^{A} &=&\{F_{kk}^{A}\},k\in \llbracket{1,6}\rrbracket.
\end{eqnarray}

(iii.1). When $n\equiv j-k\equiv 1,$ gcd$(1,6)=1$, $\frac{6}{\text{gcd}(1,6)}=6.
$ Eqs. (\ref{eq2.5},\ref{eq2.8}) yield
\begin{eqnarray}
F_{1}^{A} &=&\{\rho _{21},\rho _{32},\rho _{43},\rho _{54},\rho _{65},\rho
_{16}\}; \\
F_{(2,1)}^{A}
&=&F_{(3,2)}^{A}=F_{(4,3)}^{A}=F_{(5,4)}^{A}=F_{(6,5)}^{A}=F_{(1,6)}^{A}=\{%
\rho _{21},\rho _{32},\rho _{43},\rho _{54},\rho _{65},\rho _{16}\}.
\end{eqnarray}

(iii.2). When $n\equiv j-k\equiv 2,$ gcd$(2,6)=2$, $\frac{6}{\text{gcd}(2,6)}=3.
$ Eqs. (\ref{eq2.5},\ref{eq2.8}) yield
\begin{eqnarray}
F_{2}^{A} &=&\{\rho _{31},\rho _{42},\rho _{53},\rho _{64},\rho _{15},\rho
_{26}\}; \\
F_{(3,1)}^{A} &=&F_{(5,3)}^{A}=F_{(1,5)}^{A}=\{\rho _{31},\rho _{53},\rho
_{15}\}; \\
F_{(4,2)}^{A} &=&F_{(6,4)}^{A}=F_{(2,6)}^{A}=\{\rho _{42},\rho _{64},\rho
_{26}\}.
\end{eqnarray}

(iii.3). When $n\equiv j-k\equiv 3,$ gcd$(3,6)=3$, $\frac{6}{\text{gcd}(3,6)}=2.
$ Eqs. (\ref{eq2.5},\ref{eq2.8}) yield
\begin{eqnarray}
F_{3}^{A} &=&\{\rho _{41},\rho _{52},\rho _{63},\rho _{14},\rho _{25},\rho
_{36}\}; \\
F_{(4,1)}^{A} &=&F_{(1,4)}^{A}=\{\rho _{41},\rho _{14}\}; \\
F_{(5,2)}^{A} &=&F_{(2,5)}^{A}=\{\rho _{52},\rho _{25}\}; \\
F_{(6,3)}^{A} &=&F_{(3,6)}^{A}=\{\rho _{63},\rho _{36}\}.
\end{eqnarray}%
(iii.4). When $n\equiv j-k\equiv 4,$ gcd$(4,6)=2$, $\frac{6}{\text{gcd}(4,6)}=3.
$ Eqs. (\ref{eq2.5},\ref{eq2.8}) yield
\begin{eqnarray}
F_{4}^{A} &=&\{\rho _{51},\rho _{62},\rho _{13},\rho _{24},\rho _{35},\rho
_{46}\}; \\
F_{(5,1)}^{A} &=&F_{(3,5)}^{A}=F_{(1,3)}^{A}=\{\rho _{51},\rho _{35},\rho
_{13}\}; \\
F_{(6,2)}^{A} &=&F_{(4,6)}^{A}=F_{(2,4)}^{A}=\{\rho _{62},\rho _{46},\rho
_{24}\}.
\end{eqnarray}%
(iii.5). When $n\equiv j-k\equiv 5,$ gcd$(5,6)=1$, $\frac{6}{\text{gcd}(5,6)}=6.
$ Eqs. (\ref{eq2.5},\ref{eq2.8}) yield
\begin{eqnarray}
F_{5}^{A} &=&\{\rho _{61},\rho _{56},\rho _{45},\rho _{34},\rho _{23},\rho
_{12}\}; \\
F_{(6,1)}^{A}
&=&F_{(5,6)}^{A}=F_{(4,5)}^{A}=F_{(3,4)}^{A}=F_{(2,3)}^{A}=F_{(1,2)}^{A}=\{%
\rho _{61},\rho _{56},\rho _{45},\rho _{34},\rho _{23},\rho _{12}\}.
\end{eqnarray}

(\hyperlink{ii.1}{ii.1}) and (\hyperlink{ii.2}{ii.2}) yield that
$F_{(k,k)}^{A}\in\mathbb{R}, \ k\in\llbracket{1,6}\rrbracket;$
$F_{(k+3,k)}^{A}\in\mathbb{R}, \ k\in\llbracket{1,3}\rrbracket.$
With Eqs. (\ref{eq2.10},\ref{eq2.15}), we characterize $\text{dim}(V_{\text{KDr}})=15$ for $d=6$ in Eq. (\ref{eq3.1}). In Eq. (\ref{eq3.1}), each diagonal line expresses the elements in $F_{n}^{A}$; $F_{n}^{A}$ and $F_{-n}^{A}$ are conjugate, especially, $F_{0}^{A}$ and $F_{3}^{A}$ are real; the entries with the same lower left index are equal, for example, $_{(9)}\rho _{31}=_{(9)}\rho _{53}=_{(9)}\rho _{15}.$

$\text{dim}(V_{\text{KDr}})$ for general $d$ can be discussed similarly to Eq. (\ref{eq3.1}).

\section{Proof of $V_{\text{KDr}}=$span$_{\mathbb{R}}(\mathcal{E}_{\text{KD+}}^{\text{pure}})$ in Eq. (\ref{eq1.14})}    \label{Section-4}
\setcounter{equation}{0} \renewcommand%
\theequation{4.\arabic{equation}}

In this section we prove $V_{\text{KDr}}=$span$_{\mathbb{R}}(\mathcal{E}_{\text{KD+}}^{\text{pure}})$. Obviously, it holds that span$_{\mathbb{R}}(\mathcal{E}_{\text{KD+}}^{\text{pure}})\subseteq V_{\text{KDr}}$, thus we
only need to prove $V_{\text{KDr}}\subseteq $span$_{\mathbb{R}}(\mathcal{E}_{\text{KD+}}^{\text{pure}}).$ Suppose $F\in V_{\text{KDr}}$, according to $\dim (V_{\text{KDr}})=\gamma(d)$ and Eqs. (\ref{eq2.6}-\ref{eq2.12}), we can decompose $F^{A}$ into
\begin{eqnarray}
F^{A} &=&\sum_{k=1}^{d}F_{k,k}^{A}|a_{k}\rangle \langle
a_{k}|+\sum_{2|d;k=1}^{d/2}F_{k+d/2,k}^{A}(|a_{k+d/2}\rangle \langle
a_{k}|+h.c.)    \nonumber  \\
&&+\sum_{n\in \llbracket{1,d-1}\rrbracket,n\neq d/2,k\in \llbracket{1,\text{%
gcd}(n,d)}\rrbracket,}F_{k+\text{gcd}(n,d),k}^{A}\sum_{m=1}^{\frac{d}{\text{gcd}(n,d)}%
}|a_{k+(m+1)\text{gcd}(n,d)}\rangle \langle a_{k+m\text{gcd}(n,d)}|,  \label{eq4.1}
\end{eqnarray}%
where $F_{k,k}^{A}\in
\mathbb{R},$ $F_{k+d/2,k}^{A}\in\mathbb{R},$ $(|a_{k}\rangle \langle a_{k+d/2}|+h.c.)=(|a_{k}\rangle \langle
a_{k+d/2}|+|a_{k+d/2}\rangle \langle a_{k}|),$ $h.c.$ is the abbreviation
for "Hermitian conjugate", and there is no the term $\sum_{2|d;k=1}^{d/2}$
if $2\nshortmid d,$ $2\nshortmid d$ means that 2 is not a divisor of $d$.
From Eqs. (\ref{eq2.13},\ref{eq2.14}), we can further write the last term in Eq. (\ref{eq4.1}) as
\begin{eqnarray}
&&\sum_{n\in \llbracket{1,d-1}\rrbracket,n\neq d/2,k\in \llbracket{1,\text{%
gcd}(n,d)}\rrbracket,}F_{k+\text{gcd}(n,d),k}^{A}\sum_{m=1}^{\frac{d}{\text{gcd}(n,d)}%
}|a_{k+(m+1)\text{gcd}(n,d)}\rangle \langle a_{k+m\text{gcd}(n,d)}|    \nonumber \\
&=&\sum_{n\in \llbracket{1,\lfloor \frac{d-1}{2}\rfloor }\rrbracket,k\in \llbracket{1,\text{gcd}(n,d)}\rrbracket}[(\text{Re}F_{k+\text{gcd}(n,d),k}^{A})%
\sum_{m=1}^{\frac{d}{\text{gcd}(n,d)}}(|a_{k+(m+1)\text{gcd}(n,d)}\rangle
\langle a_{k+m\text{gcd}(n,d)}|+h.c.)    \nonumber   \\
&&\text{ \ \ \ \ \ \ \ \ \ \ \ \ \ }+i(\text{Im}F_{k+\text{gcd}%
(n,d),k}^{A})\sum_{m=1}^{\frac{d}{\text{gcd}(n,d)}}(|a_{k+(m+1)\text{gcd}%
(n,d)}\rangle \langle a_{k+m\text{gcd}(n,d)}|-h.c.)],    \label{eq4.2}
\end{eqnarray}
where $\lfloor (d-1)/2\rfloor $ is the floor function of $(d-1)/2,$ that is,
$\lfloor (d-1)/2\rfloor =(d-1)/2$ if $2|(d-1),$ $\lfloor (d-1)/2\rfloor
=(d-2)/2$ if $2|d.$

With Eqs. (\ref{eq4.1},\ref{eq4.2}), to prove $V_{\text{KDr}}=$span$_{\mathbb{R}}(\mathcal{E}_{\text{KD+}}^{\text{pure}}),$ we need to prove the following Lemma 2.
\begin{Lemma} \label{Lemma-2}
Each operator in (iv.1) to (iv.4) below belongs to span$_{%
\mathbb{R}}(\mathcal{E}_{\text{KD+}}^{\text{pure}}).$

(iv.1). $|a_{k}\rangle \langle a_{k}|$ for any $k\in \llbracket{1,d}\rrbracket.$

(iv.2). $(|a_{k+d/2}\rangle \langle a_{k}|+h.c.)$ for $2|d$ and any $k\in
\llbracket{1,d/2}\rrbracket.$

(iv.3). $\sum_{m=1}^{\frac{d}{\text{gcd}(n,d)}}(|a_{k+(m+1)\text{gcd}%
(n,d)}\rangle \langle a_{k+m\text{gcd}(n,d)}|+h.c.)$ for $n\in \llbracket{1,\lfloor (d-1)/2 \rfloor}\rrbracket,k\in \llbracket{1,\text{gcd}(n,d)}\rrbracket.$

(iv.4). $i\sum_{m=1}^{\frac{d}{\text{gcd}(n,d)}}(|a_{k+(m+1)\text{gcd}%
(n,d)}\rangle \langle a_{k+m\text{gcd}(n,d)}|-h.c.)$ for $n\in \llbracket{1,\lfloor (d-1)/2\rfloor }\rrbracket,k\in \llbracket{1,\text{gcd}(n,d)}\rrbracket.$
\end{Lemma}
\textbf{Proof of Lemma 2.}
For case (iv.1), $|a_{k}\rangle \langle a_{k}|\in $span$_{\mathbb{R}}(\mathcal{E}_{\text{KD+}}^{\text{pure}})$ obviously holds.

For cases (iv.2), (iv.3) and (iv.4), from Eq. (\ref{eq1.5}) we have
\begin{eqnarray}
|\psi _{d_{1},d_{2},j,k}\rangle \langle \psi _{d_{1},d_{2},j,k}| &=&\frac{1}{%
d_{1}}\sum_{j^{\prime },j^{\prime \prime }=1}^{d_{1}}\omega
_{d_{1}}^{j(j^{\prime }-j^{\prime \prime })}|a_{k+j^{\prime }d_{2}}\rangle
\langle a_{k+j^{\prime \prime }d_{2}}|.   \label{eq4.3}
\end{eqnarray}%

For case (iv.2), using Eq. (\ref{eq4.3}) and the fact $\omega _{2}=\exp (i\frac{%
2\pi }{2})=-1\in\mathbb{R},$ we can check that
\begin{eqnarray}
&&|a_{k+d/2}\rangle \langle a_{k}|+|a_{k}\rangle \langle a_{k+d/2}|    \nonumber   \\
&=&-|\psi _{d_{1}=2,d_{2}=d/2,j=1,k}\rangle \langle \psi
_{d_{1}=2,d_{2}=d/2,j=1,k}|+|\psi _{d_{1}=2,d_{2}=d/2,j=2,k}\rangle \langle
\psi _{d_{1}=2,d_{2}=d/2,j=2,k}|\in \text{span}_{%
\mathbb{R}}(\mathcal{E}_{\text{KD+}}^{\text{pure}}).    \label{eq4.4}
\end{eqnarray}

For cases (iv.3) and (iv.4), using Eq. (\ref{eq4.3}) and the facts $%
\sum_{m=1}^{d_{1}}\omega _{d_{1}}^{jm}=d_{1}\delta _{j,d_{1}},$ $\cos \frac{%
2\pi j}{d_{1}}=\frac{1}{2}(\omega _{d_{1}}^{j}+\omega _{d_{1}}^{-j}),$ $\sin
\frac{2\pi j}{d_{1}}=\frac{1}{2i}(\omega _{d_{1}}^{j}-\omega _{d_{1}}^{-j}),$
we obtain
\begin{eqnarray}
\sum_{j=1}^{d_{1}}\omega _{d_{1}}^{j}|\psi _{d_{1},d_{2},j,k}\rangle \langle
\psi _{d_{1},d_{2},j,k}| &=&\sum_{j^{\prime }=1}^{d_{1}}|a_{k+j^{\prime
}d_{2}}\rangle \langle a_{k+(j^{\prime }+1)d_{2}}|,     \label{eq4.5}  \\
\sum_{j=1}^{d_{1}}\omega _{d_{1}}^{-j}|\psi _{d_{1},d_{2},j,k}\rangle
\langle \psi _{d_{1},d_{2},j,k}| &=&\sum_{j^{\prime
}=1}^{d_{1}}|a_{k+j^{\prime }d_{2}}\rangle \langle a_{k+(j^{\prime
}-1)d_{2}}|,             \label{eq4.6}  \\
\sum_{j=1}^{d_{1}}(\cos \frac{2\pi j}{d_{1}})|\psi _{d_{1},d_{2},j,k}\rangle
\langle \psi _{d_{1},d_{2},j,k}| &=&\frac{1}{2}\sum_{j^{\prime
}=1}^{d_{1}}(|a_{k+j^{\prime }d_{2}}\rangle \langle a_{k+(j^{\prime
}-1)d_{2}}|+h.c.),       \label{eq4.7}   \\
\sum_{j=1}^{d_{1}}(\sin \frac{2\pi j}{d_{1}})|\psi _{d_{1},d_{2},j,k}\rangle
\langle \psi _{d_{1},d_{2},j,k}| &=&\frac{i}{2}\sum_{j^{\prime
}=1}^{d_{1}}(|a_{k+j^{\prime }d_{2}}\rangle \langle a_{k+(j^{\prime
}-1)d_{2}}|-h.c.),      \label{eq4.8}
\end{eqnarray}%
Letting $d_{2}=\gcd (n,d),$ $d_{1}=\frac{d}{\gcd (n,d)},$ we see that the
operators in cases (iv.3) and (iv.4) belong to span$_{%
\mathbb{R}}(\mathcal{E}_{\text{KD+}}^{\text{pure}}).$

We then completed the proof of Lemma 2 and completed the proof of $V_{\text{%
KDr}}=$span$_{\mathbb{R}}(\mathcal{E}_{\text{KD+}}^{\text{pure}})$ in Eq. (\ref{eq1.14}). $\hfill\blacksquare$

\section{Summary}   \label{Section-5}

We proved that in a $d-$dimensional complex Hilbert space $\boldsymbol{H}%
_{d}^{C}$ and fixed $d\times d$ DFT matrix, an operator $F$ on $\boldsymbol{H%
}_{d}^{C}$ is Hermitian and KD real if and only if $F$ can be expressed as a
real linear combination of pure KD positive states, that is, $V_{\text{KDr}}=
$span$_{\mathbb{R}}(\mathcal{E}_{\text{KD+}}^{\text{pure}}).$ We also find that the dimension
of $V_{\text{KDr}}$ is just the gcd-sum function, $\dim (V_{\text{KDr}%
})=\gamma (d).$  These results have potential applications to determine all KD positive
states, especially the conjecture that any KD positive state can be
expressed as the convex combination of pure KD states.

\section*{ACKNOWLEDGMENTS}

This work was supported by the National Natural Science Foundation of China (Grant No. 12471443). The author thanks Kailiang Lin for helpful discussions.
\hypertarget{Appendix}{}

\section*{Appendix: gcd-sum function and Proposition \ref{Proposition-1}}
\setcounter{equation}{0} \renewcommand%
\theequation{A\arabic{equation}}

For completeness of this work, in this appendix we provide some details for
the gcd-sum function $\gamma(d)$ defined in Eq. (\ref{eq1.12}) and Proposition \ref{Proposition-1}. More
details can be found for examples in Refs. \cite{Rosen-2011-book,Toth-2010-JIS}.

In number theory, an arithmetic function $f$ is a function that is defined
for all positive integers. $f$ is called multiplicative if
\begin{equation}
f(mn)=f(m)f(n) \text{\ \ if \ } \gcd (m,n)=1.  \label{eqA.1}
\end{equation}
$f$ is multiplicative if and only if
\begin{equation}
f(1)=1, \ \  f(d)=f(p_{1}^{\alpha _{1}})f(p_{2}^{\alpha _{2}})...f(p_{s}^{\alpha_{s}}),    \label{eqA.2}
\end{equation}
where $d$ is any positive integer with the standard prime factorization $%
d=p_{1}^{\alpha _{1}}p_{2}^{\alpha _{2}}...p_{s}^{\alpha _{s}}$ defined in
Eq. (\ref{eq1.11}),

The Euler phi function $\phi (d)$ is defined to be the number of positive
integers not exceeding $d$ that are relatively prime to $d,$ i.e.,
\begin{equation}
\phi (d)=|\{n:n\in \llbracket{1,d}\rrbracket,\gcd (n,d)=1\}|.    \label{eqA.3}
\end{equation}%
For the standard prime factorization of $d$ in Eq. (\ref{eq1.11}), $\phi (d)$ reads
\begin{equation}
\phi (d)=d\Pi _{j=1}^{s}\left( 1-\frac{1}{p_{j}}\right).    \label{eqA.4}
\end{equation}
$\phi (d)$ is multiplicative.

For the standard prime factorization of $d$ in Eq. (\ref{eq1.11}), suppose $m|d,$ then $%
m $ must have the standard prime factorization of the form
\begin{equation}
m=p_{1}^{\beta _{1}}p_{2}^{\beta _{2}}...p_{s}^{\beta _{s}},    \label{eqA.5}
\end{equation}
where $\beta _{j}\in \llbracket{0,\alpha _{j}}\rrbracket$ for any $j\in \llbracket{1,s}\rrbracket.$ Consequently,
\begin{eqnarray*}
\gamma (d) &=&\sum_{n=1}^{d}\gcd (n,d) \\
&=&\sum_{m\in \llbracket{1,d}\rrbracket;m|d}\left[m\sum_{n:\gcd (n,d)=m}1\right] \\
&=&\sum_{m\in \llbracket{1,d}\rrbracket;m|d}\left[m\sum_{n:\gcd (\frac{n}{m},\frac{d}{%
m})=1}1\right]\ \  \\
&=&\sum_{m\in \llbracket{1,d}\rrbracket;m|d}\left[m\phi \left(\frac{d}{m}\right)\right]\ \  \\
\ &=&\sum_{m\in \llbracket{1,d}\rrbracket;m|d}\left[\frac{d}{m}\phi (m)\right]\ \text{\ \ \
\ \ \ \ \ \ \ \ \ \ \ \ \ \ [exchanging }m\text{ and }\frac{d}{m}\text{]} \\
&=&d\sum_{\{0\leq s_{j}\leq \alpha _{j}\}_{j=1}^{r}}\frac{\phi
(p_{1}^{s_{1}}p_{2}^{s_{2}}...p_{r}^{s_{r}})\ }{%
p_{1}^{s_{1}}p_{2}^{s_{2}}...p_{r}^{s_{r}}}\text{\ \ \ \ \ \ \ \ \ \ \ \ \ \
[using Eq. (\ref{eqA.5})]} \\
&=&d\sum_{\{0\leq s_{j}\leq \alpha _{j}\}_{j=1}^{r}}\frac{\phi
(p_{1}^{s_{1}})}{p_{1}^{s_{1}}}\frac{\phi (p_{2}^{s_{2}})}{p_{2}^{s_{2}}}...%
\frac{\phi (p_{r}^{s_{r}})}{p_{r}^{s_{r}}}\text{ \ \ \ \ \ [using Eq. (\ref{eqA.2})]} \\
&=&d\left[\sum_{0\leq s_{1}\leq \alpha _{1}}\frac{\phi (p_{1}^{s_{1}})}{%
p_{1}^{s_{1}}}\right]...\left[\sum_{0\leq s_{r}\leq \alpha _{r}}\frac{\phi
(p_{r}^{s_{r}})}{p_{r}^{s_{r}}}\right]   \\
&=&d\left[1+\alpha _{1}\left(1-\frac{1}{p_{1}}\right)\right]...\left[1+\alpha _{r}\left(1-\frac{1}{p_{r}}\right)\right].%
\text{ \ \ \ \ \ [using Eq. (\ref{eqA.4})]}
\end{eqnarray*}
Eq. (\ref{eq1.12}) then follows.

For the arithmetic function $f,$ the summatory function of $f$ is defined as
\begin{equation}
F(d)=\sum_{m\in \llbracket{1,d}\rrbracket,m|d}f(m).    \label{eqA.6}
\end{equation}
The M\"{o}bius inversion formula says that Eq. (\ref{eqA.6}) yields 
\begin{equation}
f(d)=\sum_{m\in \llbracket{1,d}\rrbracket,m|d}\mu (m)F\left(\frac{d}{m}\right),    \label{eqA.7}
\end{equation}
where $\mu (d)$ is the M\"{o}bius function, for $d=p_{1}^{\alpha
_{1}}p_{2}^{\alpha _{2}}...p_{s}^{\alpha _{s}}$ defined in Eq. (\ref{eq1.11}), $\mu (d)$
reads
\begin{equation}
\mu (d)=\left\{
\begin{aligned}
&1,\ \ \ \ \ \ \ \ \ \  \text{if }d=1;\\
&(-1)^{r},\ \ \    \text{if} \ d=p_{1}p_{2}...p_{r}, \ \{p_{1},p_{2},...,p_{r}\} \ \  \text{are distinct primes};   \\
&0, \ \ \ \ \ \ \ \ \ \  \text{otherwise.}
\end{aligned}\right.        \label{eqA.8}
\end{equation}

Let $\tau (d)$ be the number of positive divisors of $d,$ for $%
d=p_{1}^{\alpha _{1}}p_{2}^{\alpha _{2}}...p_{s}^{\alpha _{s}}$ defined in
Eq. (\ref{eq1.11}), $\tau (d)$ reads
\begin{equation}
\tau (d)=\Pi _{j=1}^{s}(1+\alpha _{j}).   \label{eqA.9}
\end{equation}

We know that, if $f$ is multiplicative, then the summatory function $F$ is
multiplicative; conversely, if $F$ is multiplicative then $f$ is
multiplicative; $\mu (d)$ and $\tau (d)$ are all multiplicative. From Eqs.
(\ref{eq1.12},\ref{eqA.9},\ref{eqA.2}), it is obvious that $\gamma (d)$ and $d\tau (d)$ are all multiplicative.
Moreover, Eq. (\ref{eq2.17}) holds if and only if Eq. (\ref{eq2.18}) holds. Hence, to prove Proposition \ref{Proposition-1}, we only need to prove that Eq. (\ref{eq2.18}) holds.

We now prove Eq. (\ref{eq2.18}). Since $d\tau (d)$ is multiplicative,  then  $\gamma (d)=\sum_{m|d,m\in \llbracket{1,d}\rrbracket}m\tau (m)\mu (%
\frac{d}{m})$ is multiplicative. Eq. (\ref{eqA.2}) further implies that $\gamma (1)=1$
and $\gamma (d)=\gamma (p_{1}^{\alpha _{1}})\gamma (p_{2}^{\alpha
_{2}})...\gamma (p_{s}^{\alpha _{s}})$. Thus we only need to prove
\begin{equation}
\gamma (p_{j}^{\alpha _{j}})=p_{j}^{\alpha _{j}}\left[ 1+\alpha _{j}\left(1-\frac{%
1}{p_{j}}\right)\right] \text{ for any }\alpha _{j}\geq 1.   \label{eqA.10}
\end{equation}
Eq. (\ref{eqA.10}) holds because
\begin{eqnarray*}
\gamma (p_{j}^{\alpha _{j}}) &=&\sum_{m|p_{j}^{\alpha _{j}},m\in \llbracket{1,p_{j}^{\alpha _{j}}}\rrbracket}m\tau (m)\mu \left( \frac{p_{j}^{\alpha _{j}}}{m}\right)  \\
&=&\sum_{\beta _{1}=0}^{\alpha _{1}}p_{1}^{\beta_{1}}\tau (p_{1}^{\beta
_{1}})\mu \left( \frac{p_{1}^{\alpha _{1}}}{p_{1}^{\beta _{1}}}\right)  \text{\ \ \ \ \ \ \ \ \ \ \ \ \ \ \ \ \ [\text{letting} \ $m=p_{1}^{\beta _{1}}$]}  \\
&=&p_{1}^{\alpha _{1}}\tau (p_{1}^{\alpha _{1}})-p_{1}^{\alpha _{1}-1}\tau
(p_{1}^{\alpha _{1}-1})  \text{ \ \ \ \ \ \ \ \ \ [using Eq. (\ref{eqA.8})]}  \\
&=&p_{1}^{\alpha _{1}}(1+\alpha _{1})-p_{1}^{\alpha _{1}-1}\alpha _{1}   \text{ \ \ \ \ \ \ \ \ \ \ \ \ \ \ \ \ \ [using Eq. (\ref{eqA.9})]} \\
&=&p_{1}^{\alpha _{1}}\left[ 1+\alpha _{1}\left(1-\frac{1}{p_{j}}\right)\right].
\end{eqnarray*}
Eqs. (\ref{eqA.10},\ref{eq2.18}) and Proposition \ref{Proposition-1} then follow.


%

\end{document}